\documentclass[%
 reprint,
 amsmath,amssymb,
 aps,
]{revtex4-1}

\usepackage{graphicx}
\usepackage{dcolumn}
\usepackage{bm}
\usepackage{subfigure}
\usepackage{color}

\def\sr{Sr$_2$RuO$_4$\ }
\def\srs{Sr$_2$RuO$_4$}
\def\ml{${\bf{M}}_{\textrm{LC}}$\ }
\def\mic{$\bf{M}_{\textrm{IC}}$\ }

\def\mc{${\bf{M}_{\textrm{LC}}^{\textrm{(1)}}}$\ }

\def\mlc2{${\bf{M}_{\textrm{LC}}^{\textrm{(2)}}}$\ }
\def\mlcs{${\bf{M}_{\textrm{LC}}^{\textrm{(2)}}}$}
\newcommand{\nk}{n{\bf{k}}}

\begin{document}


\title{Modern Theory for the Orbital Moment in a Superconductor}

\author{Joshua Robbins}
\author{James F. Annett}
\author{Martin Gradhand}
\affiliation{%
 H. H. Wills Physics Laboratory, University of Bristol, Bristol BS8 1TL, United Kingdom\\
 }%

\collaboration{Bristol-Bath Centre for Doctoral Training in Condensed Matter Physics}



\date{\today}

\begin{abstract}
The chiral $p$-wave superconducting state is comprised of spin triplet Cooper pairs carrying a finite orbital angular momentum. For the case of a periodic lattice, calculating the net magnetisation arising from this orbital component presents a challenge as the circulation operator $\hat{\bf{r}} \times \hat{\bf{p}}$ is not well defined in the Bloch representation. This difficulty has been overcome in the normal state, for which a modern theory is firmly established. Here, we derive the extension of this normal state approach, generating a theory which is valid for a general superconducting state, and go on to perform model calculations for a chiral $p$-wave state in \srs . The results suggest that the magnitude of the elusive edge current in \sr is finite, but lies below experimental resolution. This provides a possible solution to the long-standing controversy concerning the gap symmetry of the superconducting state in this material.
\end{abstract}

\pacs{Valid PACS appear here}
\maketitle



An unconventional superconducting state exhibits a lower order of symmetry than the $s$-wave singlet pairing observed in conventional BCS superconductors. An example of this is the chiral $p$-wave paired state, which arises in conjunction with a breaking of time-reversal symmetry at the superconducting transition \cite{kal1}. Such a state consists of spin triplet Cooper pairs carrying a finite orbital angular momentum. The symmetry breaking associated with this pairing theoretically facilitates a number of new and exotic phenomena, such as the Kerr effect \cite{xia1,xia2} and edge currents \cite{mats,stone}.



Of major significance in the study of this class of materials is the topological nature of superconducting states with chiral symmetry \cite{qi1,sato1}. A chiral edge mode in a topological superconducting state would support a protected Majorana bound state confined to the edges of the sample \cite{sato,leij}. The existence of these bound states is inextricably linked to the orbital moment of the spin triplet Cooper pairs, as both phenomena arise from the chiral nature of the superconducting order parameter.

Given this interest, it is surprising that there currently exists no general framework with which to calculate the total orbital magnetic moment in a superconducting state. The orbital angular momentum carried by the Cooper pairs should, in principle, lead directly to an orbital magnetisation in the superconducting lattice. Contributions to the magnetic moment are expected from edge currents \cite{mats,stone}, while bulk contributions are also predicted in multi-orbital systems \cite{ann1}. The goal of this letter is to present a general approach to this problem.

A rigorous theory for the orbital magnetisation in a normal state periodic lattice has been defined previously \cite{thon1,cer1}. Obtaining a formalism of this nature had been an outstanding issue due to the problem of evaluating the circulation operator ($\hat{\bf{r}} \times \hat{\bf{p}}$) in a Bloch representation. In an infinite lattice, the position operator ($\hat{\bf{r}}$) is unbound and the cell-periodic Bloch functions ($u_{\bf{k}} (\bf{r})$) are not localised. The coexistence of these two factors means that the position expectation values of Bloch wavefunctions cannot be evaluated directly. The normal state theory was developed by reformulating the problem in a localised basis, the Wannier representation \cite{cer1,mar}. Here, we extend this formalism to the orbital magnetisation in the superconducting state.

The new theory for the orbital moment in an infinite periodic lattice has previously been applied to cases of insulators and metals, for both single-band and multi-band configurations \cite{cer1}. The derivation introduced two distinct contributions to the total moment, referred to as the ``local" and ``itinerant" circulations. The terms correspond to orbital moments generated by the movement of the centres of mass of orbital wavefunctions (itinerant), and the moment due to self-rotation about their centres of mass (local). 

Extending this theory to a general superconducting state, we obtain equivalent expressions for the local and itinerant contributions. We further break down the local contribution by performing a tight-binding expansion, extracting the purely on-site component defined previously \cite{ann1}. The formalism developed here will then be applied to a multi-band tight-binding model of \srs.

We begin our analysis by giving an outline of the derivation of the orbital moment in the superconducting state. In second quantised form, the operator for the total orbital angular momentum in an arbitrary state is given by:

\begin{equation} \label{eq1}
\hat{L}_z = \int d {\bf{r}} \, {\hat{a}}^{\dagger} \! ({\bf{r}}) \, \hat{l}_z \, {\hat{a}} ({\bf{r}})
\end{equation}

\noindent where $a^\dagger$, $a$ are Fermi creation and annihilation operators, respectively, and $\hat{l}_z = [\hat{\bf{r}} \times \hat{\bf{p}}]_{z}$. The total orbital magnetic moment is then given by $\gamma \langle \hat{L}_z \rangle$, where $\gamma = -e/(2m_e)$.

In order to obtain a second quantised operator valid for a gapped state, we perform the Bogoliubov-Valatin transformation on the creation and annihilation operators \cite{kett}:

\begin{subequations} \label{eq1a}

\begin{align} 
{\hat{a}} &= \sum_{\nk} {\theta}_{\nk} ({\bf{r}}) \hat{\gamma}_{\nk} + {\chi}_{\nk}^{*} ({\bf{r}}) \hat{\gamma}_{\nk}^{\dagger} \\
{\hat{a}}^{\dagger} &= \sum_{\nk} {\theta}_{\nk}^{*} ({\bf{r}}) \hat{\gamma}_{\nk}^{\dagger} + {\chi}_{\nk} ({\bf{r}}) \hat{\gamma}_{\nk}
\end{align}

\end{subequations}

\noindent where $n$ is the number of spin-resolved bands, ${\bf{k}}$ is the Bloch wavevector and $\gamma^{\dagger},\gamma$ are quasiparticle creation and annihilation operators. The functions  ${\theta}$, ${\chi}$ are, respectively, electron and hole components of a Bloch-type wavefunction $\psi$. This transformation recasts the equation into an expression for the orbital moment arising from Bogoliubov quasiparticles which appear as excitations in a superconductor.

To obtain the total orbital moment in an arbitrary superconducting state, we compute the expectation value of the transformed operator by applying the following relations:

\begin{subequations} \label{eq1b}

\vspace{-8pt}

\begin{align}
\langle \hat{\gamma}_{\nk}^{\dagger}  \hat{\gamma}_{n'{\bf{k}}'} \rangle &= \delta_{nn'} \delta_{{\bf{k}}{\bf{k}}'} f_{\nk} \\
\langle \hat{\gamma}_{\nk} \hat{\gamma}_{n'{\bf{k}}'}^{\dagger} \rangle &= \delta_{nn'} \delta_{{\bf{k}}{\bf{k}}'} (1-f_{\nk}) \\
\langle \hat{\gamma}_{\nk}  \hat{\gamma}_{n'{\bf{k}}'} \rangle &= \langle \hat{\gamma}_{\nk}^{\dagger}  \hat{\gamma}_{n'{\bf{k}}'}^{\dagger} \rangle = 0
\end{align}



\end{subequations}

\noindent where $f$ is the Fermi-Dirac function. The transformed equation and it's associated operators then take the following form:

\begin{equation} \label{eq2}
\langle \hat{L}_z \rangle = \sum_{n{\bf{k}}} \int d {\bf{r}} \, {\psi}_{n{\bf{k}}}^{\dagger} ({\bf{r}}) \, {\bf{{L}}}_z \, {\psi}_{n{\bf{k}}} ({\bf{r}})
\end{equation}


\begin{equation}
{\bf{{L}}}_z=\begin{pmatrix} \hat{l}_z f_{n{\bf{k}}} & 0 \\ 0 & -\hat{l}_{z}^{*} (1-f_{n{\bf{k}}}) \end{pmatrix} ,  \, {\psi}_{n{\bf{k}}} ({\bf{r}})= \begin{pmatrix} {\theta}_{n{\bf{k}}} ({\bf{r}}) \\ {\chi}_{n{\bf{k}}} ({\bf{r}}) \end{pmatrix} \nonumber
\end{equation}


At this point, we can defer to the derivation laid out for the normal state in terms of Wannier orbitals \cite{thon1,cer1}, where we now consider two-component Wannier wavefunctions containing electron and hole amplitudes in correspondence with the Bloch-type eigenfunctions. We also introduce the cell-periodic components of the Bloch wavefunctions, $(u_{n{\bf{k}}} ({\bf{r}}),v_{n{\bf{k}}} ({\bf{r}}))={\textrm{e}}^{-i {\bf{k}}\cdot{\bf{r}}}({\theta}_{n{\bf{k}}} ({\bf{r}}),{\chi}_{n{\bf{k}}} ({\bf{r}}))$.

Following the steps of this derivation, we are able to remove the dependence of Eq. (\ref{eq2}) on the problematic operators $\hat{\bf{r}}$ and $\hat{\bf{v}}$. Performing a Fourier transform on the real space expressions obtained via this approach, we obtain two reciprocal space expressions which generate the orbital magnetisation via Brillouin zone integrals:

\begin{widetext}
\begin{equation} \label{eq4}
{\bf{M}}_{\textrm{LC}}=-\gamma \textrm{Im}\left[  \int_{BZ} \frac{d {\bf{k}}}{(2 \pi)^3} \sum_{n} \Big( \langle \partial_{\bf{k}} u_{n{\bf{k}}} | \times \hat{H}_{\bf{k}} | \partial_{\bf{k}} u_{n{\bf{k}}} \rangle f_{n{\bf{k}}} - \langle \partial_{\bf{k}} v_{n{\bf{k}}} | \times \hat{H}_{\bf{k}}^{*} | \partial_{\bf{k}} v_{n{\bf{k}}} \rangle (1-f_{n{\bf{k}}}) \Big) \right]
\end{equation}

\begin{equation} \label{eq5}
{\bf{M}}_{\textrm{IC}}=\gamma \textrm{Im}\left[  \int_{BZ} \frac{d {\bf{k}}}{(2 \pi)^3} \sum_{n} E_{n{\bf{k}}} \Big( \langle \partial_{\bf{k}} u_{n{\bf{k}}} | \times  | \partial_{\bf{k}} u_{n{\bf{k}}} \rangle f_{n{\bf{k}}} + \langle \partial_{\bf{k}} v_{n{\bf{k}}} | \times | \partial_{\bf{k}} v_{n{\bf{k}}} \rangle (1-f_{n{\bf{k}}}) \Big) \right]
\end{equation}
\end{widetext}

\noindent where LC and IC refer to local and itinerant circulations, as defined previously \cite{thon1}, and the total magnetisation is given by ${\bf{M}} = {\bf{M}}_{\textrm{LC}} + {\bf{M}}_{\textrm{IC}}$. We have divided by the unit cell volume, to convert from the magnetic moment to magnetisation, and also introduced Dirac notation where, crucially, the expectation values taken in equations (\ref{eq4}) and (\ref{eq5}) are now evaluated for the unit cell only. These equations constitute our central result: a comprehensive framework for computing the total orbital magnetisation in a general bulk superconducting state.

The cell-periodic functions are obtained through self-consistent calculation of the Bogoliubov-de Gennes (BdG) equation:

\begin{equation} \label{eq3}
\begin{pmatrix} \hat{H}_{\bf{k}} ({\bf{r}}) & \Delta ({\bf{r}}) \\ \Delta^{\dagger} ({\bf{r}}) & -\hat{H}_{-\bf{k}}^{*} ({\bf{r}}) \end{pmatrix} \begin{pmatrix} u_{n{\bf{k}}} ({\bf{r}}) \\ v_{n{\bf{k}}} ({\bf{r}}) \end{pmatrix} = E_{n{\bf{k}}} \begin{pmatrix} u_{n{\bf{k}}} ({\bf{r}}) \\  v_{n{\bf{k}}} ({\bf{r}}) \end{pmatrix}
\end{equation}

\noindent where $\hat{H}_{\bf{k}}$ is the ${\bf{k}}$-dependent normal state Hamiltonian \cite{grad1}. The gap function ($\Delta$) enforces the symmetry of the superconducting state in question.

In order to perform model calculations, we must recast the Bloch equations into a tight-binding representation. Performing the $k$-derivatives in (\ref{eq4}) and (\ref{eq5}) and expanding in terms of the Bloch wavefunctions, we obtain:

\begin{subequations} \label{eq9}

\begin{align} 
\partial_{\bf{k}} u_{n{\bf{k}}} ({\bf{r}}) &= \textrm{e}^{-i {\bf{k}} \cdot {\bf{r}}} \big(\partial_{\bf{k}} \theta_{n{\bf{k}}} ({\bf{r}}) - i {\bf{r}} \theta_{n{\bf{k}}} ({\bf{r}}) \big) \\ \partial_{\bf{k}} v_{n{\bf{k}}} ({\bf{r}}) &= \textrm{e}^{-i {\bf{k}} \cdot {\bf{r}}} \big(\partial_{\bf{k}} \chi_{n{\bf{k}}} ({\bf{r}}) - i {\bf{r}} \chi_{n{\bf{k}}} ({\bf{r}}) \big)
\end{align}

\end{subequations}

Substituting Eqs. (\ref{eq9}) into (\ref{eq5}), we find one term containing ${\bf{r}} \times {\bf{r}}$, which will vanish. For the local component, however, this does not occur and we can split the equation into two parts of the form $\partial_{\bf{k}} \theta_{n{\bf{k}}}^{*} \times \hat{H} \partial_{\bf{k}} \theta_{n{\bf{k}}}$ and $\theta_{n{\bf{k}}}^{*} [ {\bf{r}} \times \hat{H} {\bf{r}} ] \theta_{n{\bf{k}}}$ respectively. Using the standard definition of the velocity operator, ${\bf{r}} \times \hat{H} {\bf{r}}$ can be re-written as $-i \hat{l}_z$.

Having re-written Eq. (\ref{eq5}) in terms of $\theta$, $\chi$ we can subsequently apply a general tight-binding expansion of the Bloch wavefunction via:


\begin{equation} \label{eq12}
\begin{pmatrix}
{\theta}_{n{\bf{k}}} ({\bf{r}}) \\ {\chi}_{n{\bf{k}}} ({\bf{r}}) \end{pmatrix} = \sum_{L,{\bf{R}}} \textrm{e}^{i {\bf{k}} \cdot {\bf{R}}} \begin{pmatrix} u_{nL} ({\bf{k}}) \\ v_{nL} ({\bf{k}}) \end{pmatrix} \phi_L ({\bf{r}}-{\bf{R}})
\end{equation}

\vspace{2pt}

\noindent where $L$ is the orbital index and $\phi_L$ is the corresponding orbital wavefunction. Substituting Eq. (\ref{eq12}) into (\ref{eq4}), we obtain the following terms:

\begin{widetext}
\begin{equation} \label{eq10}
{\bf{M}}_{\textrm{LC}}^{\textrm{(1)}}=-\gamma \textrm{Im} \bigg[\sum_{nLL'} \int_{BZ} \frac{d {\bf{k}}}{(2 \pi)^3} \Big( \partial_{\bf{k}} u_{nL}^{*}({\bf{k}}) \times \hat{H}_{LL'}({\bf{k}}) \partial_{\bf{k}} u_{nL'}({\bf{k}}) f_{n{\bf{k}}} - \partial_{\bf{k}} v_{nL}^{*}({\bf{k}}) \times \hat{H}_{LL'}^{*}({\bf{k}}) \partial_{\bf{k}} v_{nL'}({\bf{k}}) (1-f_{n{\bf{k}}}) \Big) \bigg]
\end{equation}

\begin{equation} \label{eq11}
{\bf{M}}_{\textrm{LC}}^{\textrm{(2)}}=\gamma \textrm{Re} \Bigg[\sum_{nLL'} \int_{BZ} \frac{d {\bf{k}}}{(2 \pi)^3} \Big( u_{nL}^{*}({\bf{k}}) \left( \hat{l}_{z,LL'} \right) u_{nL'}({\bf{k}}) f_{n{\bf{k}}} + v_{nL}^{*}({\bf{k}}) \left( \hat{l}_{z,LL'}^{*} \right) v_{nL'}({\bf{k}}) (1-f_{n{\bf{k}}}) \Big) \Bigg]
\end{equation}
\end{widetext}

The eigenvectors ($u_{nL}$, $v_{nL}$) are computed by solving Eq. (\ref{eq3}) self-consistently in the tight-binding basis. The terms $\hat{H}_{LL'}$ represent the matrix elements of the tight-binding Hamiltonian. Similarly, the matrix elements $\hat{l}_{z,LL'}$ correspond to the orbital angular momentum expectation values of the orbitals contained in the tight-binding basis. These elements can be calculated by direct consideration of the spherical harmonics of the basis.

The second term, \mlcs , is identical to the purely on-site orbital moment computed previously \cite{ann1}. We therefore label \mlc2 as the ``on-site" component, and continue to refer to \mc as the local contribution.



Now that the framework for calculating the magnetic moment has been set up, we briefly outline the model for \sr that will be used to perform the calculations. The superconducting state of \sr is widely believed to exhibit chiral $p$-wave superconductivity below its transition temperature of 1.5 K \cite{rev1,rev2}, such that the superconducting order parameter is given by ${\bf{d}} \! \sim \! (\sin k_x \pm i \sin k_y){\bf{\hat{z}}}$. This hypothesis is supported by measurements of spin susceptibility \cite{ish2,duf1} and indirect observations of time-reversal symmetry breaking at T$_c$ \cite{luke1}. In addition, a finite Kerr shift has been measured in this material \cite{xia1}, providing direct evidence of a macroscopic orbital magnetisation in the bulk superconducting state. 

The classification of \sr as a $p$-wave superconductor remains a point of controversy, however, as phenomenological and quasiclassical approaches have predicted that large edge currents should accompany the single-band chiral superconducting state \cite{sig1,mats,hua}. Such currents have remained elusive despite years of intensive experimental work \cite{kirt,hicks1,curr}. A large surface-based current would provide a significant contribution to the total orbital magnetisation. By generating a full theoretical description of the orbital magnetic moment and its various sources in such a state, we provide a vital avenue through which we can attempt to reconcile these observations with theory.

We have constructed a three-dimensional tight-binding Hamiltonian consisting of three Ru 4$d$ orbitals ($d_{xy}$, $d_{xz}$ and $d_{yz}$)  contributing to the normal state Fermi surface, resulting in a 2D band (denoted $\gamma$) and two quasi-1D bands ($\alpha$ and $\beta$). In many approaches to modelling \srs , the model is formulated such that superconductivity arises primarily on $\gamma$, with accompanying gaps on $\alpha$ and $\beta$ arising only through proximity effects. Here, we treat all three bands on an equal footing, resulting in a fully multi-band superconductivity picture. The included 1D bands exhibit horizontal line nodes, leading to the experimentally observed power law for the specific heat. This model has been covered in more detail previously \cite{ann2,grad1}.

\begin{figure}[t]
\includegraphics[scale=1]{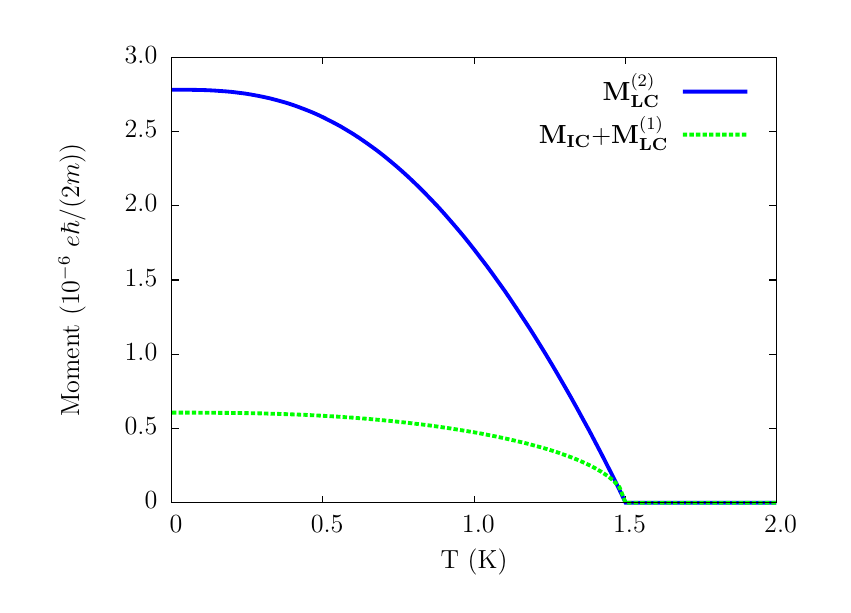}
\caption{On-site moment \mlc2 alongside the sum of the itinerant and local components \mic + \mc for the unperturbed model.}
\label{fig1}
\end{figure}

The distinct contributions to the magnetisation in this model are plotted in Fig. \ref{fig1}. It should be noted that the contributions \mic and \mc diverge to plus and minus infinity respectively as $T$ approaches $T_c$. This problem arises due to the fact that these components are not separately gauge-invariant, and thus we must take the sum of the two. It has been shown previously that gauge-invariant forms of the normal state equations for \ml and \mic can be obtained \cite{sou1}. However, this requires an absolute distinction between the occupied and unoccupied states in the electron bandstructure. The Bogoliubov transformation enforces mixing of the electron and hole states. This mixing is essential to recover the quasiparticle bandstructure of the superconducting state, but prevents any attempts to project excitations onto occupied states and thus our expressions cannot be converted into separately gauge-invariant forms.

Comparison of the itinerant contributions and the on-site component reveal that they are of similar orders of magnitude. This corresponds to a magnetic field of the order of $\mu$G, which is below the resolution of the most recent attempts to experimentally identify an edge current in \sr via magnetometry measurements ($\sim 2.5$ mG \cite{curr}). We can therefore surmise that it is possible the edge currents accompanying the chiral state in \sr have remained elusive because their magnitude is significantly smaller than has been previously hypothesised. 

The reason for this suppression in the orbital moment in comparison to other theoretical approaches likely lies in the multi-band, nodal nature of our tight-binding model and gap structure. Significantly, this result agrees with other experimental and theoretical observations which support the idea that multi-band superconductivity is prevalent in this material. It has been shown previously that inter-orbital transitions are necessary in order for the Kerr effect to arise intrinsically in the superconducting state \cite{grad1,robb1}. In order to see the effect in a single band picture, extrinsic mechanisms such as skew scattering must be considered \cite{gor2}. The inclusion of the additional 1D, line nodal bands also leads to the correct specific heat below $T_c$ \cite{robb1}. The nodeless 2D band would not produce the experimentally observed power laws in heat capacity \cite{nish} or NMR spin relaxation rate \cite{ish3}.

We also wish to assess the influence of spin-orbit coupling (SOC) on the magnetic moment in the chiral state. To do this, we compare results using a tight-binding model with an additional spin-orbit Hamiltonian derived in an on-site approximation. As was shown previously \cite{robb1}, a model including spin-orbit coupling with coupling parameter $\lambda=12.5$ meV is able to replicate experimental features such as the Fermi surface, bandwidth and heat capacity. In the following, we compare the non-SOC case ($\lambda=0$) to the case with SOC ($\lambda=12.5$ meV).

\begin{figure}[t]
\includegraphics[scale=1]{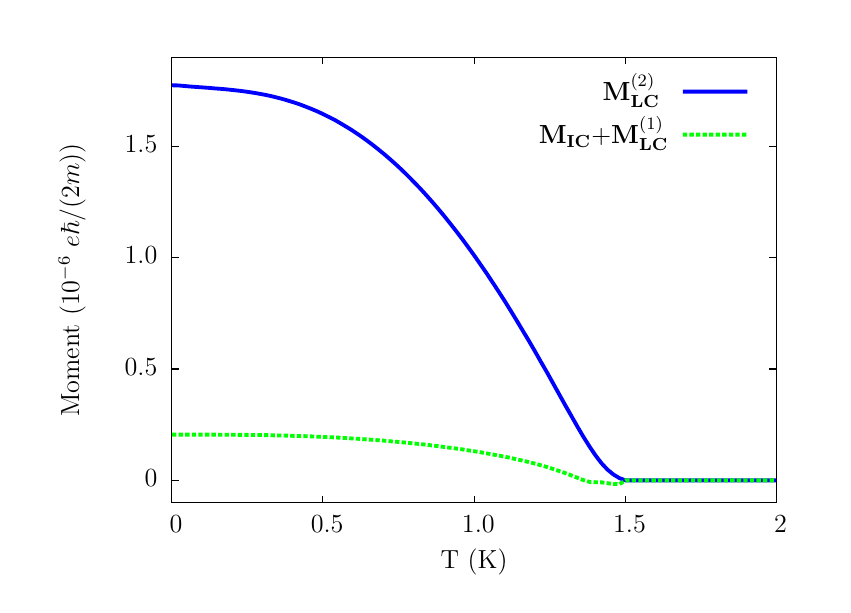}
\caption{On-site moment \mlc2 alongside the sum of the itinerant and local components \mic + \mc for the model including spin-orbit coupling.}
\label{fig4}
\end{figure}

\begin{figure}[t]
\includegraphics[scale=1]{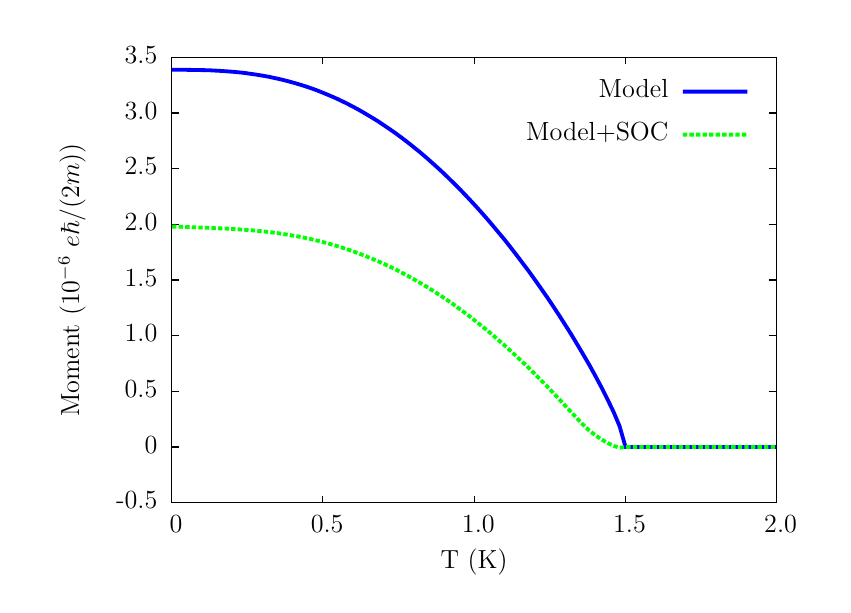}
\caption{Total orbital magnetic moment for the model with and without SOC.}
\label{fig2a}
\end{figure}

The equivalent plots for the model including SOC are displayed in Fig. \ref{fig4}. We also compare the total magnetisation for the cases with and without SOC (see Fig. \ref{fig2a}), noting that the addition of SOC results in a clear suppression of the total orbital moment. We observe a significant quantitative reduction in all contributions, without any qualitative differences in the temperature-dependence displayed. This suppression is also of similar order to that seen in the Kerr effect under the influence of SOC as reported previously \cite{robb1}.

In order to fully assess the influence of SOC, it is informative to also compute the spin moment of the chiral state. To do this we start with the equation for the spin expectation value in the orbital basis:

\begin{equation} \label{eq6}
\langle \hat{S}_z \rangle = \sum_{mm'\sigma\sigma'} \langle m \sigma | \frac{\hbar}{2} {\mathbf{\sigma_z}} | m' \sigma' \rangle n_{mm'}^{\sigma\sigma'}
\end{equation}

\noindent where $m,\sigma$ are the orbital and spin degrees of freedom respectively and $n$ are the single particle density matrices. 

The density matrix can be evaluated in terms of solutions to the BdG equation, while the $\sigma_z$ matrix elements are $\pm 1$ for $\sigma=\sigma'=\pm 1$ and $m=m'$. The final expression is then:

\begin{equation} \label{eq7}
\langle \hat{S}_z \rangle = \sum_{m} \frac{\hbar}{2} \left( n_{mm}^{\uparrow\uparrow} - n_{mm}^{\downarrow\downarrow} \right)
\end{equation}

\begin{equation} \label{eq8}
n_{mm}^{\sigma\sigma}=\frac{1}{N} \sum_{n\bf{k}} |u_{n{\bf{k}}}^{\sigma}|^2 f(E_{n{\bf{k}}}) + |v_{n{\bf{k}}}^{\sigma}|^2 \left(1-f(E_{n{\bf{k}}})\right)
\end{equation}

\begin{figure}[t]
\includegraphics[scale=1]{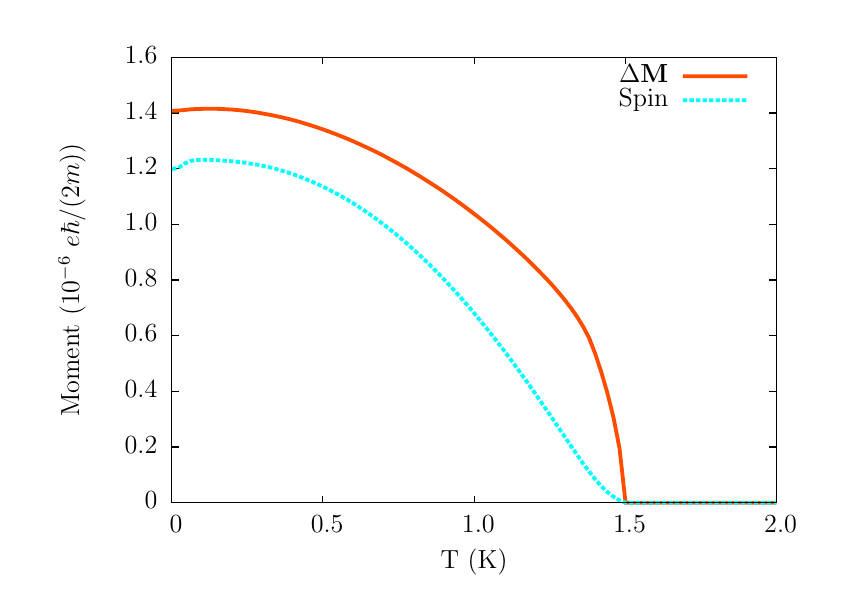}
\caption{Spin moment in the model including SOC alongside the difference in orbital moment between the two models.}
\label{fig2}
\end{figure}

\noindent The spin magnetic moment is given by $\gamma_s \langle \hat{S}_z \rangle$ where $\gamma_s = -e g/(2 m_e)$ and $g$ is the spin gyromagnetic ratio.

It is interesting to note here that the spin moment in this context becomes non-zero when SOC is included (see Fig. \ref{fig2}), but is zero otherwise. The spin moment in the SOC regime is of similar order to the reduction in the total orbital moment induced by the spin-orbit interaction (which we have denoted $\Delta {\bf{M}}$). This would suggest that the spin-orbit interaction mediates a transfer of magnetic moment from the orbital degrees of freedom (where it arises from the chiral order parameter) to the spin degrees (which are otherwise disordered).

This observation provides an interesting insight into the origin of the Kerr effect, a phenomenon which is driven by the anomalous Hall conductivity present in systems with a finite orbital moment. The microscopic origin of this effect in unconventional superconductors has been extensively debated \cite{gor2,min2}. The current controversy concerns whether the origin is extrinsic (i.e. arising from disorder \cite{kim1,lut2,gor1}) or an intrinsic mechanism arising from coupling of the pair state to orbital degrees of freedom at the Fermi level \cite{wyso1,grad1,robb1}. 

In the normal state ferromagnet, the intrinsic mechanism facilitating the Kerr effect is induced by coupling of the ordered spins to the orbital component via SOC. Namely, the symmetry breaking in the spin degree of freedom is transferred to the orbital component via the spin-orbit interaction. This is a clear analogue to the results reported here, where orbital order arises naturally due to the chiral superconducting order parameter, and is then reduced via coupling to the disordered spin component. These results coincide with the observations reported previously, where the magnitude of the Kerr shift in the same chiral superconducting model was also shown to be suppressed by a similar order following the introduction of SOC \cite{robb1}. Our model is thus able to effectively describe an intrinsic origin of the anomalous phenomena observed in \srs.

\begin{figure}[h]
\flushleft
\subfigure{
  \includegraphics[scale=0.496]{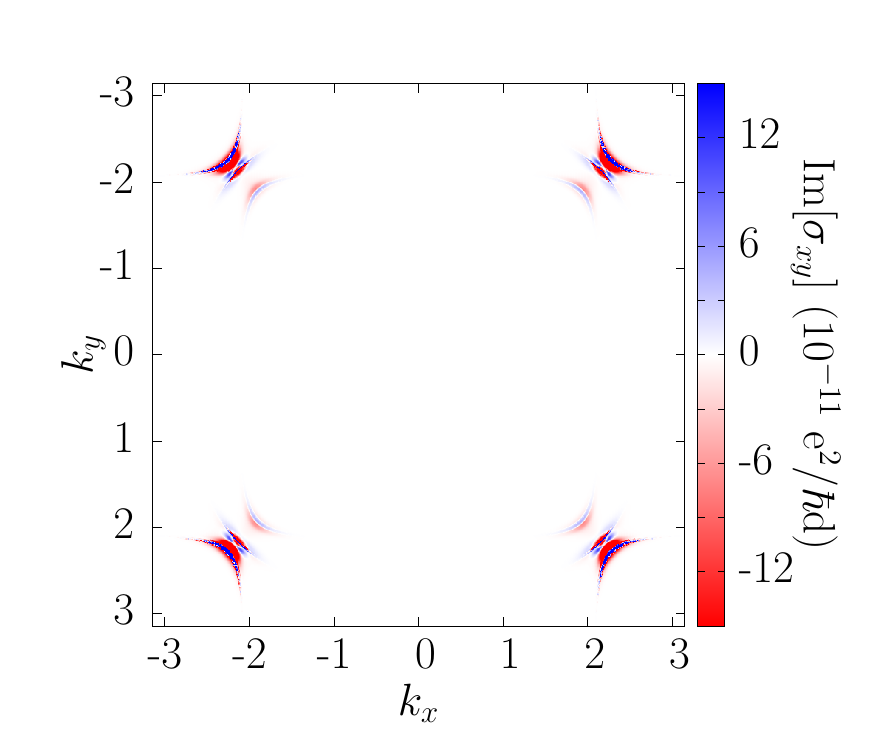}}
\hspace{-17pt}	
\subfigure{
  \includegraphics[scale=0.496]{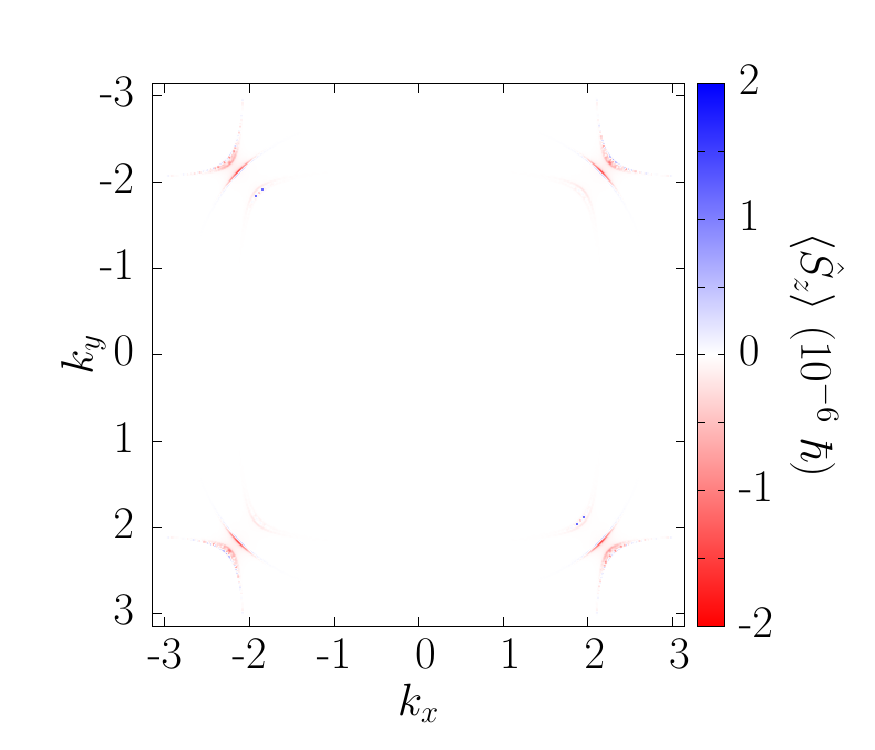}}
 \caption{a) Berry curvature contributions in the Brillouin zone integrated along $k_z$, $T$=0, with spin-orbit coupling. b) $k_x - k_y$ resolved plot of the spin moment in the Brillouin zone. The $k_z$ dependence has been integrated out.}
\label{fig3}
\end{figure}

This analysis of the influence of SOC is further supported by assessing the regions of the Brillouin zone in which the spin moment arises (see Fig. \ref{fig3}). We see here that the spin moment is present in regions of near-degeneracies between the orbital degrees of freedom in the bandstructure. These regions on the Brillouin zone contribute strongly to the Berry curvature, which gives rise to an anomalous Hall conductivity \cite{grad2}. This implies that these regions contain the highest density of ordered orbital moments, which in turn suggests that the spin magnetisation is arising directly as a result of coupling of the spins to the orbital degree of freedom.


In conclusion, a new formalism for computing the orbital magnetisation in a superconductor has been derived and calculations for the model chiral $p$-wave superconductor \sr have been performed. The results suggest that early estimations of the itinerant magnetisation in this state were too generous and that the magnitude of edge currents may lie well below the resolution of magnetometry based investigations. This same model has been shown to also give a physically reasonable estimate of the observed Kerr effect \cite{grad1}. An interesting insight into the influence of SOC on a magnetic superconducting state has also been highlighted.

It should be stressed that the general result here is not restricted to the model used. We note that our theory would also apply to other pairing states which have been proposed for \srs, such as the chiral $d$-wave \cite{zut1}, $f$-wave \cite{has1}, or long range $p$-wave states \cite{sca1}. In addition, the equations presented here could be used to investigate the unconventional pairing symmetries observed in other materials, such as the underdoped cuprates and heavy-fermion compounds.

\bibliographystyle{unsrt}
\bibliography{orb}

\end{document}